\documentclass[twocolumn,showpacs,preprintnumbers]{revtex4}%
\usepackage{graphicx}
\usepackage{dcolumn}
\usepackage{bm}%
\usepackage{amsmath}%
\usepackage{amsfonts}%
\usepackage{amssymb}

\newcommand{\be}{\begin{equation}}
\newcommand{\ee}{\end{equation}}
\newcommand{\ba}{\begin{eqnarray}}
\newcommand{\ea}{\end{eqnarray}}

\begin{document}
\title{ $g_{s}$-factor quenching with quantum statistics}
\author{S. Hamieh}
\email{hamieh@KVI.nl}
\affiliation{\textit{Kernfysisch Versneller
Instituut, Zernikelaan 25, 9747 AA Groningen, The Netherlands.}}
\altaffiliation{University of Nijmegen, NL 6525 ED Nijmegen, The
Netherlands.}
\author{C. Bertulani}
\email{bertulani@physics.arizona.edu}
\affiliation{\textit{Department of Physics, University of Arizona,
Tucson, Arizona 85721, USA.}}
\date{\today}

\begin{abstract}
We discuss the effect of quantum statistics on the $g_{s}$-factor
quenching in nuclear medium. We have found that the two nucleon
correlation function can be described by a Gaussian with correlation
length $\sigma\sim1.2$ fm. An estimate of the M1 quenching strength
is given. The effect of quantum statistics on the retardation of the
paramagnetic susceptibility is also discussed.

\end{abstract}
\maketitle

Quenching of the spin-M1 transition strength and the $g_{s}$-factor
in nuclear medium has been reported in different experiments and has
been found, in general, to be more pronounced for heavy nuclei.
 The Sakai group
\cite{saki1} found that the Gamow Teller (GT) transition (analogous
to M1) is located at 30 MeV with $\sim93\%$ sum rule value. Thus,
this cannot be explained by $\Delta$-isobar as the GT transition
should be at 300 MeV. A possible explanation of this puzzle can be
found in \cite{akit1}. Core polarization or $p-h$ mixing with higher
order contributions is the most promising mechanism but it seems to
fail for heavy nuclei \cite{akit1}. All these effects have been well
studied in the literature and one may naturally ask if new phenomena
could contribute to the quenching of M1 strength. This is our goal
in this paper. We propose a simple mechanism, namely, the effect of
quantum statistics on the nucleon spin correlation and consequently
on the transition strength. To the best of our knowledge, for
unknown reasons, this effect has been rarely studied explicitly in
nuclear physics \cite{bert1}.

In a macroscopic system, quantum statistics has led to a number of interesting
phenomena such as ferromagnetism, superconductivity and superfluidity.
Recently, it has been shown experimentally that the free magnetic moments in
solids that should manifest themselves in terms of Curie laws, are replaced by
power laws for a variety of materials, $e.g.$ salt $\mathrm{LiHo_{x}Y_{1-x}%
F_{4}}$. This effect is attributed to quantum correlations of spins in the
Heisenberg chain, as seen in Fig. 1 of \cite{ghos1}. Motivated by these
results, the aim of this paper is to evaluate the spin correlations in nuclear
matter due to quantum statistics and its effects on the $g_{s}$-factor. It
should be understood that the spin correlations discussed here are purely
quantum phenomena  due to the indistinguishability of nucleons  without
explicit reference to the nuclear interaction \cite{6}.

This paper is organized as follows. In the next section we  evaluate the spin
correlation function: the quantum and the classical correlation between two
protons inside a nucleus. In section II we discuss the effects of the spin
correlations on the quenched $g_{s}$-factor in calculations of the M1
strength. Our conclusion is given in section III.

\section{\textit{Spin correlations in nuclear matter}}

\begin{figure}[ptb]
\includegraphics{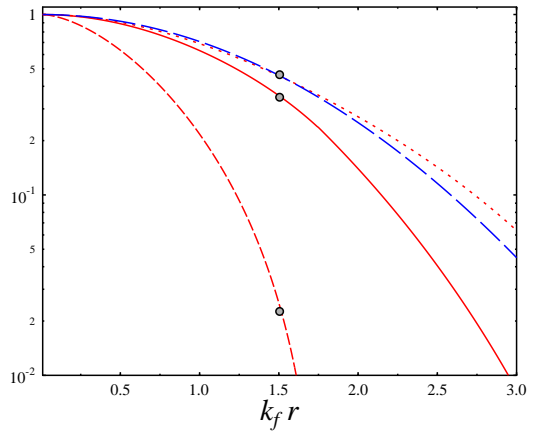} \vspace{-0.7cm}\caption{The correlation function
at $0^{\circ}$ (solid line), the relative entropy of entanglement
(dashed line) and the classical correlation function (dotted line),
are shown as a function of $k_{f}r$. Small circles represent our
predictions for the nuclear matter where we have used
$k_{f}\sim1.27\,\mathrm{fm}^{-1}$ and $r\sim1.2$ fm. The large
dashed line represents the Gaussian approximation of the mixing
parameter.}%
\label{werner}%
\end{figure}

To illustrate the correlations between nucleons, the Fermi gas model is used
\cite{101,100}. Thus the ground state of the system is described by
\begin{equation}
|\Psi_{0}\rangle=\prod_{|k|\leq k_{f}}a_{\sigma\mathbf{k}}^{\dag}|0\rangle\,,
\end{equation}
where $|0\rangle$ is the vacuum state and in the continuum limits
$[a_{\sigma\mathbf{k}}^{\dag},a_{\sigma^{\prime}\mathbf{k^{\prime}}}%
]=\delta_{\sigma\sigma^{\prime}}\delta(\mathbf{k-k^{\prime}})$.

In coordinate representation, the density matrix elements of two nucleons are
\begin{equation}
\langle\Psi_{0}|\psi^{\dag}_{\sigma_{2}^{\prime} \mathbf{r_{2}^{\prime}}}%
\psi^{\dag}_{\sigma_{1}^{\prime} \mathbf{r_{1}^{\prime}}}\psi_{\sigma_{2}
\mathbf{r_{2}}}\psi_{\sigma_{1} \mathbf{r_{1}}}|\Psi_{0}\rangle\,,
\end{equation}
where $\psi_{\sigma\mathbf{r}}=\frac{1}{V}\sum_{\mathbf{k}}e^{i\mathbf{k\cdot
r}}a_{\sigma\mathbf{k}}$.

A straightforward calculation yields the explicit form of the two-particle
space-spin density matrix \cite{99}
\begin{align}
\rho^{sp}  & =\frac{1}{2}\left(  g(\mathbf{r_{1}-r_{1}^{\prime}}%
)g(\mathbf{r_{2}-r_{2}^{\prime}})\delta_{\sigma_{1}\sigma_{1}^{\prime}}%
\delta_{\sigma_{2}\sigma_{2}^{\prime}}\right.  \nonumber\\
& \left.  -g(\mathbf{r_{1}-r_{2}^{\prime}})g(\mathbf{r_{2}-r_{1}^{\prime}%
})\delta_{\sigma_{1}\sigma_{2}^{\prime}}\delta_{\sigma_{2}\sigma_{1}^{\prime}%
}\right)  \,
\end{align}
with $g(\mathbf{r})=\frac{1}{V}\sum_{\mathbf{k}}e^{i\mathbf{k\cdot r}}$.

Depending on the space density matrix, two spins may be entangled.

The two-spin density matrix, depending on the relative distance
between two nucleons, $r=|\mathbf{r_{1}-r_{2}}|$, reads
\[
\rho^{s}=\frac{1}{4-2f^{2}}\left(
\begin{array}
[c]{cccc}%
1-f^{2} & 0 & 0 & 0\\
0 & 1 & -f^{2} & 0\\
0 & -f^{2} & 1 & 0\\
0 & 0 & 0 & 1-f^{2}%
\end{array}
\right)  \,,
\]
where $f(r)=3j_{1}(k_{f}r)/k_{f}r$ and $j_{1}$ is the spherical
Bessel function. We find that $\rho^{s}$ is a Werner state
characterized by a single parameter $p=f^{2}/(2-f^{2})$. This result
can be explained intuitively. In fact, because $S_{z}$ is conserved,
the density matrix should interpolate between the completely random
state and the singlet state.

In Fig.~\ref{werner} we show the correlation function \cite{ham4}
\begin{equation}
P(\theta)={\frac{N_{++}+N_{--}-N_{+-}-N_{-+}}{N_{total}}}=p\cos(\theta)\,,
\end{equation}
as function of $k_{f}r$, and the relative entropy of
entanglement~\cite{Vedr97}%
\begin{equation}
E_{r}=\min_{\rho^{\ast}\in\mathcal{D}}S(\rho\Vert\rho^{\ast}%
)\,,\label{eq:ccorr}%
\end{equation}
where $\mathcal{D}$ is a set of all separable states in the Hilbert space in
which $\rho$ is defined. Also shown in Fig.~\ref{werner} is the classical
correlation~\cite{Hami03},
\begin{equation}
\Psi(\rho)=S(\rho\Vert\rho_{A}\otimes\rho_{B})-\min_{\rho^{\ast}\in
\mathcal{D}}S(\rho\Vert\rho^{\ast})\,,\label{eq:qcorr}%
\end{equation}
where $\rho_{A}$ and $\rho_{B}$ are the reduced density matrices.
The small circles in Fig.~\ref{werner} represent our predictions for
the nuclear matter with typical values of
$k_{f}\sim1.27\,\mathrm{fm}^{-1}$ and $r\sim1.2$ fm \cite{100}. We
observe that the mixing factor, $p(r)$, can be approximated with a
Gaussian function
\begin{equation}
p(r)=e^{-r^{2}/2\sigma^{2}}\,,
\end{equation}
with $\sigma=1.2$ fm, see Fig. 1. The mixing parameter, $p(r)$, can
be measured experimentally. In fact, following Bertulani's work
\cite{bert1} for weakly bound nucleons, as in the
$\mathrm{^{1}H(^{6}He,^{2}He)^{5}H}$ reaction, one can argue that
the relative energy distribution of the two outgoing
protons from $\mathrm{^{2}He}$ is given by%
\begin{equation}
\sigma(E_{pp})\propto\sqrt{E_{pp}}\left(  1+e^{-R^{2}/2\sigma^{2}}[\cos
(\sqrt{m_{n}E_{pp}/\hbar^{2}}R)]\right)  \,,
\end{equation}
where $E_{pp}$ is the relative energy between the two protons and
$R$ is the size of the source region. Therefore, from experimental
data \cite{kors1} one can extract the mixing parameter $\sigma$
\cite{hami3}. Also our mixing parameter can be extracted from
Coulomb dissociation of $\mathrm{^{11}Li}$ \cite{simo1} (see
forthcoming paper \cite{hami3}).

\section{\textit{Quenching of $g_{s}$ and paramagnetic susceptibility}}

From the above discussion it is clear that there are non-negligible
spin quantum correlations effects in a nucleus. Therefore, \ a
natural question arises, namely, do these correlations play a role
on the observables that depend on the spin of the nucleon pair? We
start our discussion by a calculation of this effect on the M1
transition strength. Evidently, we expect that when two nucleons are
in a singlet spin state the M1 transition is suppressed, or even
forbidden, due to Pauli blocking and the energy level structure.
Similar arguments were given in \cite{moni1} upon requiring that the
recoiling nucleon lies outside the Fermi sphere. The amount of
singlet state is controlled by the factor $p(r)$, thus the number of
the suppressed transitions should be proportional to $p(r)$
\footnote{Note that the state should be entangled, {\it e.g. } 
$p(r)>1/3$ (see Fig. 1) in order to have such effect on the M1
transition}.

\begin{figure}[ptb]
\includegraphics{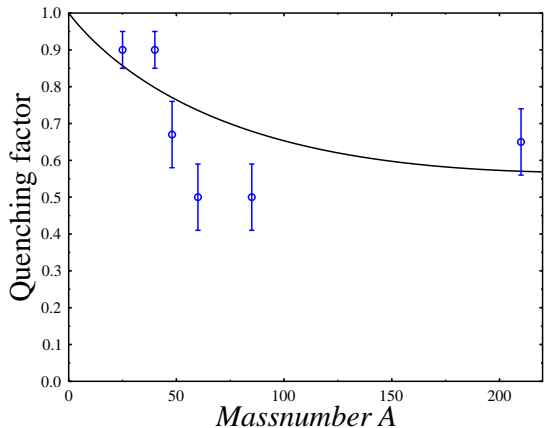} \vspace{-0.7cm}\caption{ Quenching as a function of
the mass  number, with $\gamma=0.015$.}%
\label{test}%
\end{figure}

However, we should not forget that the same correlation exists for
all nucleons. This puts another constraint on the M1 transition and
thus we need to evaluate the multiparticle correlation function.
Unfortunately this correlation function is not known \footnote{Note
that the technique used in \cite{bert1} under the assumption of a
pure state will note work in our case because the multiparticle spin
state will always be mixed}. Nevertheless, we can intuitively assume
that the relevant correlation will be two-body correlations and that
they will be distributed uniformly among all nucleons. Thus, we
expect that the $g_{s}$ quenching will be of the form
\begin{equation}
g_{s}^{\mathrm{qs}}=1-\gamma A\exp(\beta A^{2/3})\,,\label{1}%
\end{equation}
where $\gamma$ and $\beta$ are the parameters of our model. Note
that other global effects can be taken into account by multiplying
$g_{s}^{\mathrm{qs}}$
by an overall factor as follows: $g_{s}^{\mathrm{eff}}=g_{s}^{\mathrm{qs}%
}g_{s}^{\mathrm{other}}\,.$ The experimental quenching factor for M1
\cite{knup1} and our predictions are depicted in Fig. \ref{test}. The
agreement is reasonable. The fitting parameters of Eq. \ref{1} are
$\gamma=0.008\pm0.002$ and $\beta=-0.04\pm0.008$ with $\chi^{2}/\mathrm{dof}%
\sim2.6$.

This clearly gives a reasonable value for the nuclear susceptibility
\cite{knup1}
$$\chi_{\mathrm{para}}={e\over 2m}\sum_{n}{|\langle0|\sum_{i}g_{l}^{i}\mathbf{l_{i}
}+g_{s}^{i}\mathbf{s_{i}}|n\rangle|^{2}\over (E_{n}-E_{0})}\, .$$
The key to matching the experimental susceptibility is to use the
correct $g_{s}^{\mathrm{eff}},$ together with the reduced $g_{l}$
due to pion exchange \cite{hami3}.

\section{Conclusions}

We have shown in this paper that quantum statistics play a major
role in nucleonic spin correlations inside the nucleus. We have
found that observables involving the nucleon spins, \textit{e.g.} M1
transition amplitudes and/or the paramagnetic susceptibility, are
highly sensitive to spin correlations. A decisive test of our
predictions can be made \textit{e.g.} with two-proton radioactivity.
We face two major problems in our model, namely:

\begin{itemize}
\item a complete knowledge of the total density matrix and,

\item a measure of multiparticle correlations.
\end{itemize}

We hope to address these issues in the near future.

\section*{Acknowledgments}

S.H. acknowledge stimulating discussions with H. Wortche.
This work was performed as part of the research program of the
\textsl{Stichting voor Fundamenteel Onderzoek der Materie (FOM)}
with financial support from the \textsl{Nederlandse Organisatie voor
Wetenschappelijk Onderzoek } and was supported by the U.\thinspace
S.\ Department of Energy under grant No. DE-FG02-04ER41338.


\begin{thebibliography}{99}
\bibitem {100} The shell model and the optical potential model suggest that, to first
approximation, we may picture the nucleus as a potential well with
42 MeV depth filled up to about 8 MeV below the edge. Thus, $k_{f}\sim1.27\,\mathrm{fm}%
^{-1}$ and $r\sim1.2$ fm.
\bibitem {101}Similar calculations can be done in the shell model.
\end{thebibliography}
\end{document}